# High-Velocity Magnetic Domain Wall Motion Driven by Acoustic Spin Transfer Torque

**Jiacheng Lu[1], Fa Chen[1], Yiming Shu[1], Yukang Wen[1], Hang Zou[1], Yuhao Liu[1], Shiheng Liang[2], Wei Luo[1*], Yue Zhang[1,3*]**

[1]School of Integrated Circuits, Huazhong University of Science and Technology, Wuhan, 430074, China

[2]School of Physics, Hubei University, Wuhan 430062, China

[3]Songshan Lake Materials Laboratory, Dongguan, Guangdong 523808, China

*Corresponding author: luowei@hust.edu.cn (Wei Luo); yue-zhang@hust.edu.cn (Yue Zhang)

We predict high-velocity magnetic domain wall (DW) motion driven by out-of-plane acoustic spin in surface acoustic waves (SAWs). We demonstrate that the SAW propagating at a 30-degree angle relative to the *x*-axis of a 128° Y-$LiNbO_3$ substrate exhibits uniform out-of-plane spin angular momentum. This acoustic spin triggers the DW motion at a velocity exceeding 50 m/s in a way that is similar to the spin-transfer-torque (STT) effect. This phenomenon highlights the potential of acoustic spin in enabling rapid DW displacement, offering an innovative approach to developing energy-efficient spintronic devices.

SAWs and their interactions with magnetism, particularly through magnetoelastic coupling, have emerged as a pivotal research focus in the field of magnetoacoustic phenomena **[1-9]**. By inducing strain in magnetic materials, SAWs influence magnetization dynamics, paving the way for novel applications in low-dissipation spintronic and acoustic devices.

In recent years, significant progress has been made in utilizing SAWs to manipulate magnetic DWs through magnetoelastic coupling **[10-14]**. However, the velocity of DW motion induced by conventional SAWs, such as Rayleigh-type SAW, has been limited to around 1 m/s **[12]**. Simultaneously, advancements in understanding the angular momentum properties of acoustic waves have opened new possibilities for enhancing magnetoacoustic interactions. Traditionally characterized by their linear momentum, acoustic waves are now recognized for their intrinsic angular momentum, particularly spin angular momentum **[15-17]**. This has sparked growing interest in harnessing the spin angular momentum of SAWs to control magnetic moments, providing novel mechanisms for manipulating magnetic dynamics.

Our recent experiments have demonstrated when SAWs propagate at a 30-degree angle relative to the *x*-axis of a 128° Y-$LiNbO_3$ substrate, out-of-plane spin angular momentum emerges, which gives rise to nonreciprocal magnetoacoustic waves resulting from the breaking of crystal inversion symmetry **[18]**. In this work, we propose that this out-of-plane acoustic spin could induce the DW motion at a velocity higher than 50 m/s in a way similar to the STT effect, marking potential advancement for energy-efficient spintronic devices driven by SAWs.

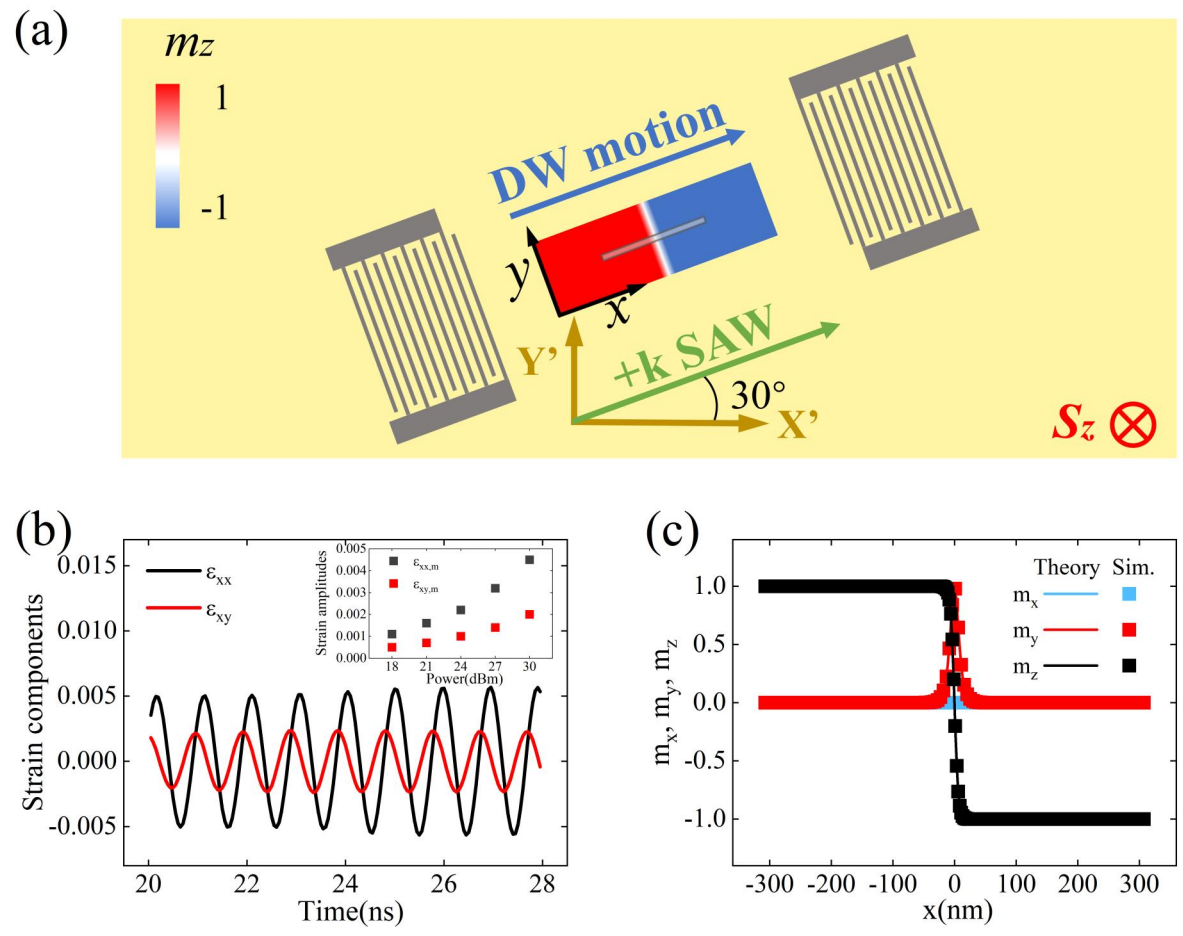

**Figure 1. (a) Schematic for the excitation of the SAW propagating at a 30-degree angle relative to the *x*-axis of a 128° Y-$LiNbO_3$ substrate and the driving of DW by this SAW, (b) Simulated $\varepsilon_{xx}$ and $\varepsilon_{xy}$ components of the strain for the 30-degree SAW at a frequency of 1.04 GHz (Inset: the variation of the amplitudes of $\varepsilon_{xx}$ and $\varepsilon_{xy}$ of the 30-degree SAW under different excitation powers), (c) Walker profile of the Bloch DW in the magnetic film with perpendicular magnetic anisotropy.**

We consider the model composed by the interdigitated transducers (IDTs) on the surface of lithium niobate (128° Y-$LiNbO_3$) with the orientation forming an angle with respect to the *x*-axis crystal axis of $LiNbO_3$ [**Figure 1(a)**]. Based on the Multiphysics simulation, we simulated the strain components of the SAW at the central frequency of 1.04 GHz under the power of 30 dBm, which is experimentally feasible. We found that when this angle is set to 30 degrees (We name it 30-degree SAW.), the strain components of the SAW are composed by $\varepsilon_{xx}$, $\varepsilon_{xy}$, $\varepsilon_{xz}$, $\varepsilon_{yz}$ and $\varepsilon_{zz}$. Here $\varepsilon_{xx}$, $\varepsilon_{xz}$, $\varepsilon_{yz}$ and $\varepsilon_{zz}$ consists of the Rayleigh SAW, which, as will be confirmed below, do not contribute to the fast DW motion. Nevertheless, the $\varepsilon_{xx}$ and $\varepsilon_{xy}$ components [**Figure 1(b)**] can generate acoustic wave spin ($\boldsymbol{S}_z$ ) along the out-of-plane orientation of the film, and the effective magnetoelastic field of this acoustic spin drives the Bloch-type DW [**Figure 1(c)**] into rapid motion.

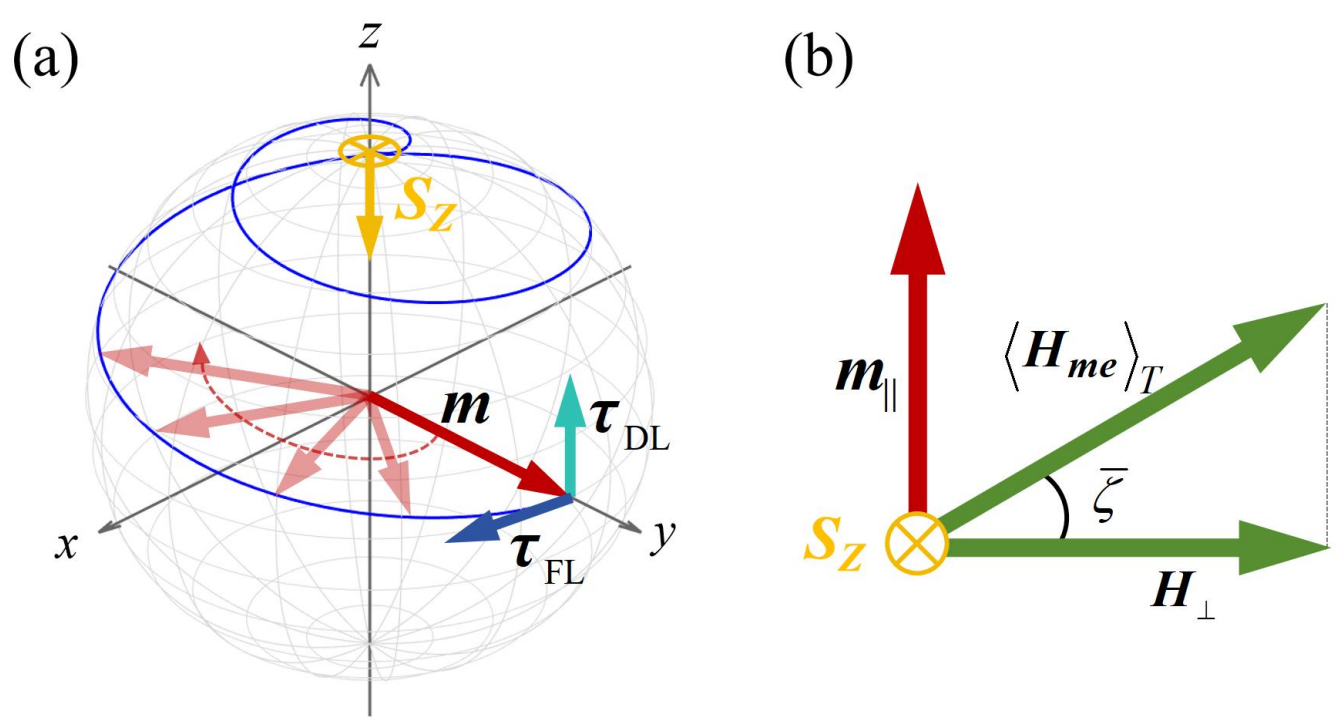

**Figure 2. (a) Schematic of the magnetization rotation in the central of DW (the red arrow) driven by the field-like (FL) and damp-like (DL) torque of the 30-degree SAW spin (the yellow arrow). The yellow arrow represents the SAW spin. (b) Schematic of the SAW spin, the average magnetoelastic field, and the magnetization in the central of DW.**

The simulated strain components for the 30-degree SAW are mainly composed by $\varepsilon_{xx}$ and $\varepsilon_{xy}$ as $\varepsilon_{xx} = Ak\sin(\omega t - kx)$ and $\varepsilon_{xy} = 1/2Ak\sin(\omega t - kx + \pi/2)$. Here $A$, $k$, and $\omega$ are the displacement amplitude, the wavenumber, and the angular frequency of the SAW. Based on $\varepsilon_{xx} = \frac{\partial u_x}{\partial x}$, $\varepsilon_{xy} = \frac{1}{2}\left(\frac{\partial u_y}{\partial x} + \frac{\partial u_x}{\partial y}\right)$, and $\boldsymbol{v} = \mathrm{d}\boldsymbol{u}/\mathrm{d}t$, the displacement $\boldsymbol{u}$ and the velocity field $\boldsymbol{v}$ was expressed as $\boldsymbol{u} = [A\cos(kx - \omega t), A\cos(kx - \omega t - \pi/2), 0]$ and $\boldsymbol{v} = [A\omega\cos(kx - \omega t - \pi/2), A\omega\cos(kx - \omega t - \pi), 0]$. The acoustic spin density $\boldsymbol{s}$ can be computed from the expression **[19]**:

$$s = \rho(u \times v). \tag{1}$$

The acoustic spin density of the 30-degree SAW was $\boldsymbol{s} = -A^2\omega\rho\boldsymbol{e}_z$. It exhibits a uniform spin density, with the spin polarization aligned along the out-of-plane direction of the magnetic film. The acoustic wave spin, in this context, refers to the rotation of the local velocity field of particles **[17]**. For the 30-degree SAW, all particles undergo coherent rotation within the *xy* plane, resulting in a uniform spin angular momentum. In addition to that, the SAW also carries orbital angular momentum, which, however, does not contribute significantly to the fast DW motion **(S1 in the Supplementary Materials).**

First, we discuss the physical mechanism for the influence of the SAW spin on the DW motion. Based on the relationship between the SAW spin and the $\varepsilon_{xx}$ and $\varepsilon_{xy}$ strain components, we have found that the magnetoelastic energy density $F_{me} = b\left(\varepsilon_{xx}m_x^2 + 2\varepsilon_{xy}m_xm_y\right)$ can be converted to $F_{me} = -\mu_0\boldsymbol{H_{me}}\cdot\boldsymbol{m}$ with effective magnetoelastic field satisfying **(S2 in the Supplementary Materials)**

$$\boldsymbol{H_{me}} = \frac{|b|k^2}{\mu_0 M_s \varepsilon_m \omega\rho}\eta\mathbf{R}(\zeta)\cdot(\boldsymbol{s}\times\boldsymbol{m}). \tag{2}$$

In Eq. (2), $\eta = \sqrt{\frac{3}{2} - \frac{\cos(2\omega t)}{2} + 2\sin(C)\sin(\omega t)}$ is the factor depicting that the strength of the magnetoelastic field oscillates around an average value $\langle H_{me}\rangle_t = \sqrt{\frac{3}{2}}\frac{|b|\varepsilon_m\sin\theta}{\mu_0 M_s}$. $\mathbf{R}(\zeta) = \begin{bmatrix}\cos\zeta & -\sin\zeta & 0\\ \sin\zeta & \cos\zeta & 0\\ 0 & 0 & 1\end{bmatrix}$ with $\zeta = \frac{\pi}{2} - \arccos\left[\frac{-b\left[\sin(\omega t) + \sin C\right]}{|b|\sqrt{\frac{3}{2} - \frac{\cos(2\omega t)}{2} + 2\sin C\sin(\omega t)}}\right]$, which indicates the counterclockwise rotation of the magnetoelastic field from $\boldsymbol{s}\times\boldsymbol{m}$ by an angle $\zeta$, which oscillates around the average value $\bar{\zeta} = \frac{\pi}{2} - \arccos\left[-\sqrt{\frac{2}{3}}\frac{b\sin C}{|b|}\right]$. The following calculation will show that the oscillations of $\boldsymbol{H_{me}}$ and $\zeta$ lead to weak fluctuation on the trajectory of the DW magnetization. Therefore, we only considered the average magnetoelastic

field expressed as:

$$\langle \boldsymbol{H}_{me} \rangle_T = \sqrt{\frac{3}{2}} \frac{|b| k^2}{\mu_0 M_s \varepsilon_m \omega \rho} \mathbf{R}\left(\bar{\zeta}\right) \cdot (\boldsymbol{s} \times \boldsymbol{m}) . \tag{3}$$

The $\boldsymbol{H}_{me}$ can be decomposed into two components that are parallel and perpendicular ( $\boldsymbol{H}_{\perp}$ ) to the magnetization. The dynamics of the magnetization in the DW was determined by the torque of $\boldsymbol{H}_{\perp}$ acting on magnetization, which is usually expressed as $\tau_1 = \boldsymbol{m} \times \boldsymbol{H}_{\perp}$ and $\tau_2 = \boldsymbol{m} \times (\boldsymbol{m} \times \boldsymbol{H}_{\perp})$ . Based on Eq. (3), it can be concluded that $\tau_1 \sim \boldsymbol{m} \times (s \times \boldsymbol{m})$ and $\tau_2 \sim s \times \boldsymbol{m}$ , which are the same to the DL and FL torque for the STT generating by an out-of-plane spin, respectively.

Unlike the STT for the spin of electrons, the DL and FL torques of acoustic spin have the same strength since they share the same coefficient in Eq. (3). Therefore, the coexistence both torques drags the rotation of magnetization towards the acoustic spin along the *z* axis, which is accompanied with continuous magnetization precession in the *xy* plane. This process sequentially happens for all the magnetization within the DW plane, resulting in the unidirectional DW motion accompanied by continuous magnetization rotation at the DW center.

To verify the above assumption, we simulated the DW motion by using the MUMAX3 simulation with the code of magnetoelastic interaction based on the parameters for perpendicularly magnetized Pt/Co and the $LiNbO_3$ substrate **[20]**: Gilbert damping coefficient $\alpha$ = 0.3, gyromagnetic ratio $\gamma = 1.76 \times 10^{11}$ rad/T·s, vacuum permeability $\mu_0 = 4\pi \times 10^{-7}$ H/m, saturation magnetization $M_S = 5.8 \times 10^5$ A/m, exchange stiffness constant $A_{ex} = 1.5 \times 10^{-11}$ J/m, and uniaxial anisotropy constant $K_u = 6 \times 10^5 \text{J/m}^3$. The velocity of SAW $v_{SAW}$ = 4000 m/s, magnetoelastic coupling coefficient $b_1 = b_2 = -15.75 \times 10^6$ J/m$^3$, and density of Co $\rho$ = 8900 kg/m$^3$.

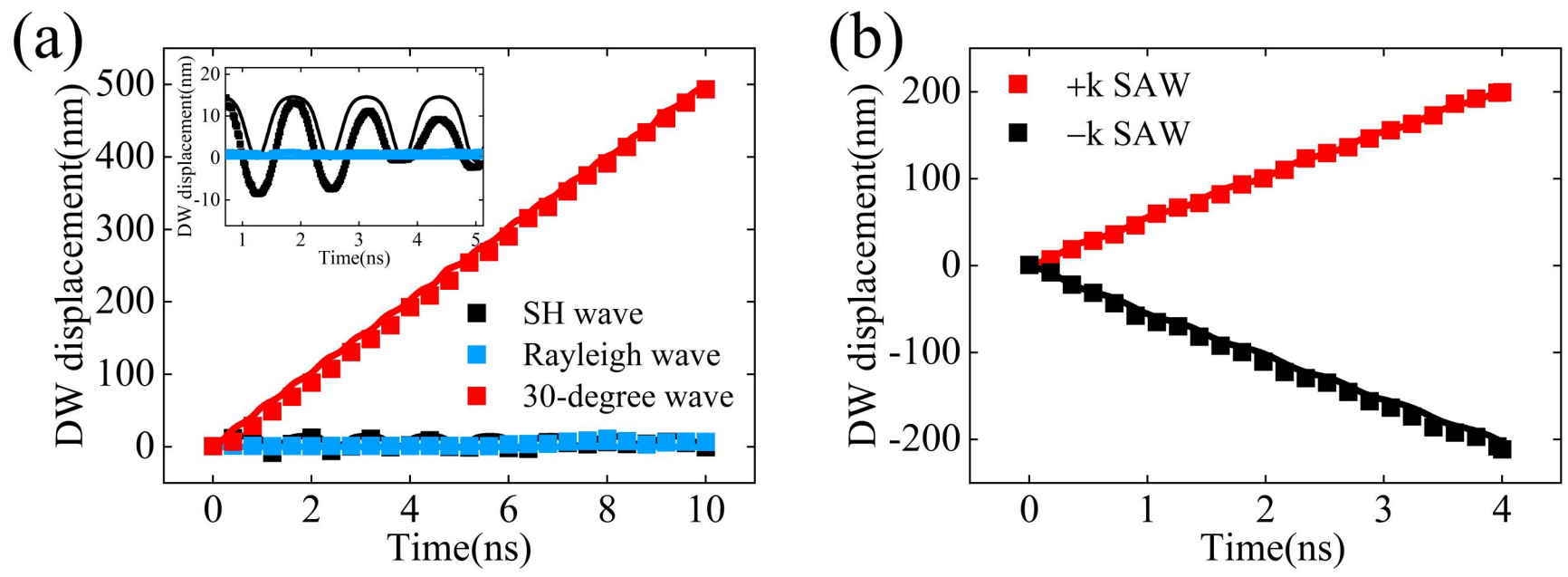

**Figure 3. (a) Comparison of the DW displacement driven by 30-degree, Rayleigh, and SH SAW at 0.8-GHz frequency (Inset: enlarged portion for the DW motion driven by the Rayleigh and SH SAW), (b) DW displacement under forward and backward propagation of the 30-degree SAW. The solid lines represent the calculation results of the CCM.**

**Figure 3(a)** demonstrates that the strain field of the 30-degree SAW induces continuous DW motion for approximately 500 nm within 10 ns. In contrast, both the Rayleigh and SH waves only result in weak oscillation, showing nearly no net displacement [**inset of Figure 3(a)**]. The influence of polarization direction of the SAW spin on DW motion was exhibited in **Figure 3(b)**. When the SAW propagates along the positive (negative) *x*-axis direction, it possesses spin angular momentum along the negative (positive) *z*-axis direction, driving the DW along the positive (negative) *x*-axis direction at a velocity of 49.98 m/s (52.80 m/s).

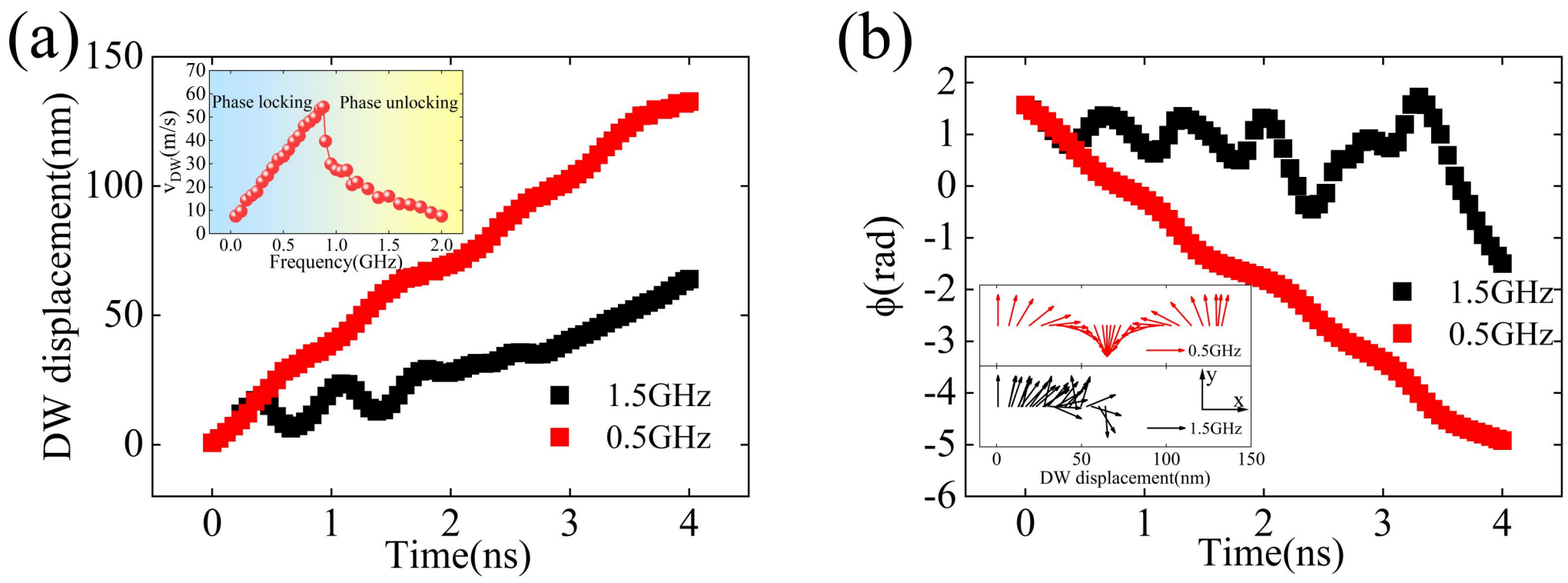

**Figure 4. (a) Displacement of the DW driven by the 30-degree SAW at the frequency of 0.5 and 1.5 GHz (Inset: Variation of DW velocity as a function of the frequency of the 30-degree SAW), (b) Variation of the azimuthal angle at the frequency of 0.5 and 1.5 GHz (Inset: Variation of the orientation of the magnetization in the DW central with the DW motion).**

**Figures 4** presents the relationship between DW motion and the acoustic wave frequency (*f*). In **Figure 4(a)**, the DW velocity initially increased with frequency, reaching a maximum value at approximately 0.88 GHz, above which it decreases sharply. This indicates the existence of an optimal frequency ($f_m$) for driving the DW. The inset in **Figure 4(b)** illustrated the precession of magnetization in the DW center at frequencies lower and higher than $f_m$, and the phase-locking (phase-unlocking) precession occurred when $f < (>) f_m$.

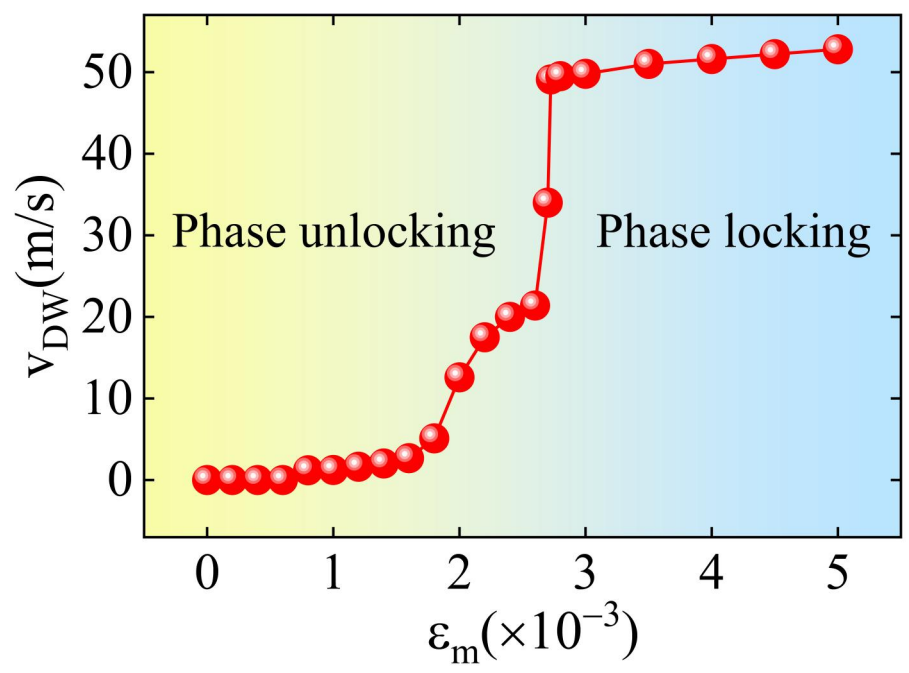

**Figure 5. DW velocity as a function of the strain amplitude for the 30-degree SAW at a frequency of 0.8 GHz.**

The amplitude of strain ($\varepsilon_m$) also influences the DW velocity in a similar way (**Figure 5**). In the simulation, above a threshold $\varepsilon_m$ around 0.0027, the DW velocity becomes almost stable, and below it, the DW velocity decreases with the reducing $\varepsilon_m$. We will show that the appearance of this threshold $\varepsilon_m$ is also relevant to the transition between the phase-locking and phase-unlocking magnetization precession.

The simulated DW motion with phase-locking characteristics can be derived by using the Collective-Coordinate-Method (CCM). We started by the Lagrangian density function **[21]**:

$$\begin{aligned} L_{dens} &= E_{dens} + \frac{M_s}{\gamma}\dot{\phi}\cos\theta \\ &= \frac{A_{ex}}{\Delta^2}\sin^2\theta + K_u\sin^2\theta + \frac{\mu_0}{2}M_s(N_x\sin^2\theta\cos^2\phi + N_y\sin^2\theta\sin^2\phi + N_z\cos^2\phi) \\ &+ b_1\varepsilon_{xx}\cos^2\phi\sin^2\theta + 2b_2\varepsilon_{xy}\sin\phi\cos\phi\sin^2\theta + \frac{M_s}{\gamma}\dot{\phi}\cos\theta. \end{aligned} \tag{4}$$

Here $\theta$, $\phi$, and $\Delta$ are the polar angle, azimuthal angle, and the DW width, and $\theta$

satisfies $\tan\frac{\theta}{2}=\exp[\frac{x-q}{\Delta}]$ with $q$ the DW central position. $N_x$, $N_y$, and $N_z$ are the demagnetizing factors in the $x$, $y$, and $z$ directions, respectively. $\frac{M_s}{\gamma}\dot{\phi}\cos\theta$ is the precession term. For simplicity, we neglect the spatial dependence of strain as the SAW wavelength is approximately 4 μm, which is much longer than the DW width, i.e., $\varepsilon_{xx} = \varepsilon_{xx,m}\sin(\omega t)$, $\varepsilon_{xy} = \varepsilon_{xy,m}\sin(\omega t + \pi/2)$. Additionally, we did not consider the influence of the variation of strain on the DW width, since we verified that the oscillation of DW width under strain of the 30-degree SAW does not make considerable impact on the DW velocity **(S3 in the Supplementary Materials)**.

The Rayleigh dissipation density function is:

$$F_{dens}=\frac{\alpha M_s}{2\gamma}\left(\frac{d\vec{m}}{dt}\right)^2=\frac{\alpha M_s}{2\gamma}\left(\frac{\dot{q}^2}{\Delta^2}+\dot{\phi}^2\right)\sin^2\theta\,. \tag{5}$$

The spatial integral of $L_{dens}$ and $F_{dens}$ yields the Lagrangian $L$ and Rayleigh dissipation function $F$. Based on the Euler-Lagrange-Rayleigh equation： $\frac{\partial L}{\partial q_i}-\frac{d}{dt}\left(\frac{\partial L}{\partial \dot{q}_i}\right)+\frac{\partial F}{\partial \dot{q}_i}=0$ with $q_i = q, \phi$, we deduced the Thiele equations:

$$\dot{\phi}+\frac{\alpha\dot{q}}{\Delta}=0\,, \tag{6}$$

$$\frac{\dot{q}}{\Delta}-\alpha\dot{\phi}=\frac{1}{2}\mu_0\gamma M_s(N_y-N_x)\sin 2\phi-\frac{\varepsilon_{xx,m}\gamma b_1}{M_s}\sin(\omega t)\sin 2\phi+\frac{2\varepsilon_{xy,m}\gamma b_2}{M_s}\cos(\omega t)\cos 2\phi\,. \tag{7}$$

The initial DW width extracted from the simulation is $\Delta$ = 5.9 nm. According to the method in Ref. **[22]**, for a square film system, $N_x \approx N_y$ . Therefore, in the following analysis, the contribution of the demagnetization term is neglected.

The crucial role of the SAW spin in driving DWs lies in two magnetoelastic terms with π/2 phase difference on the right side of Eq. (5). by defining $\varepsilon_m$ as $\varepsilon_{xx,m} = 2\varepsilon_{xy,m} = \varepsilon_m$ and $b_1 = b_2 = b$, and substituting Eq. (4) into Eq. (5), we converted the dynamics equation of $\phi$ to:

$$\left(\alpha+\frac{1}{\alpha}\right)\dot{\phi}=-\frac{\varepsilon_m\gamma b}{M_S}\cos(\omega t+2\phi)\,. \tag{8}$$

Both of the numerical solution of Eq. (8) and the simulation illustrate an approximately linear variation of $\phi$ at lower frequencies, which, as indicated in Eq. (4), should lead to constant DW velocity. On the other hand, the linear variation of $\phi$ will further give rise to constant $\omega t + 2\phi$ in Eq. (8) and we denote this constant as $C$. Therefore, $\dot{\phi}=-\frac{\omega}{2}$, demonstrating the phase-locking magnetization precession in a moving DW at the velocity:

$$\dot{q}=\frac{\omega\Delta}{2\alpha}=\frac{\Delta\varepsilon_m\gamma b}{\left(\alpha^2+1\right)M_S}\cos\left(C\right). \tag{9}$$

The threshold values of frequency and $\varepsilon_m$ in **Figure 4** and **Figure 5** originate from the issue that $|\cos(C)|$ should not exceed 1. Based on Eq. (9), the linear increase of the DW velocity cannot be kept when the frequency exceeds the threshold:

$$\omega_{th}=\frac{2\alpha\varepsilon_m\gamma\,|\,b\,|}{\left(\alpha^2+1\right)M_s}, \tag{10}$$

or when the $\varepsilon_m$ is smaller than the critical value as:

$$\varepsilon_{m,th}=\frac{\left(\alpha^2+1\right)M_s}{2\alpha\gamma\,|\,b\,|}\omega, \tag{11}$$

and the magnetization precession in DW will become phase-unlocking. This gives rise to the oscillation of the DW displacement and the gradual reducing of the DW velocity. The estimation of the limit of angular frequency is in good agreement with the simulation: Under the strain amplitude of 0.003, the theoretical threshold angular frequency is $\omega_{th}$ = 2π × 1.26 GHz, close to the simulated $\omega_{th}$ (2π × 0.88 GHz). On the other hand, under the acoustic wave frequency of 0.8 GHz, the theoretical $\varepsilon_{m,th}$ is 0.002, which is also close to the simulated value $\varepsilon_{m,th}$ = 0.0027. The discrepancy between the theoretical model results and the simulation may stem from the assumed constant DW width in the CCM. In principle, when the strain exceeds the saturation value, the DW velocity would stabilize. However, the simulation shows slight increase of the DW velocity when the strain exceeds 0.0027. This is mainly because of the modulation of the DW width under SAW **(S3 in the Supplementary Materials)**.

The transition between phase-locking and phase-unlocking magnetization precession also can be understood based on the variation of the angle $\zeta$ in Fig. 2(b). During the phase-locking process, $C$ represents a constant for the magnetization at the DW center, and the angle $\zeta$ decreases with increasing frequency or decreasing strain amplitude. At the threshold frequency or strain amplitude, $C = \pi$, and the average $\boldsymbol{H}_{me}$ becomes strictly perpendicular to the magnetization. Under this situation, the strength of the field for driving the magnetization precession reaches its maximum value, and the DW moves at its highest velocity.

For comparison, we also conducted brief discussion on the interaction between DWs and pure Rayleigh or SH SAWs **(S4 in the Supplementary Materials)**. In the Thiele equation, either type of the SAW contributes to opposite precession at the same angular velocity, which leads to opposite DW motion at the same velocity strength and is insufficient to enable unidirectional DW motion.

In conclusion, we have demonstrated that SAWs maintain a uniform constant spin angular momentum as they propagate at a 30-degree angle relative to the $x$-axis crystal orientation of 128° Y-$LiNbO_3$. We have confirmed that the out-of-plane acoustic spin offers the FL and DL torques that have the same expression as that of STT. These torques induces DW motion at velocities exceeding 50 m/s, significantly surpassing the velocities achieved with conventional Rayleigh and SH SAWs. In addition, the DW motion undergoes a transition between phase-locking and phase-unlocking regimes at critical values of SAW frequency or strain amplitude. These findings present a novel mechanism for achieving high-speed DW motion by utilizing the acoustic spin, paving the way for energy-efficient spintronic device applications.

## Acknowledgements

This work was supported by the National Key Research and Development Program of China (Grant No. 2022YFE0103300) and the open research fund of Songshan Lake Materials Laboratory

(Grant No. 2023SLABFN26).